\documentclass[aps,prl,twocolumn,superscriptaddress,longbibliography]{revtex4-2}

\usepackage{graphicx}
\usepackage{amssymb,amsmath}
\usepackage{bm,bbold}
\usepackage{dcolumn}
\usepackage{float}
\usepackage[OT1]{fontenc} 
\usepackage{url}
\usepackage{mathrsfs}
\usepackage{slashed,comment}
\usepackage{color}
\usepackage{verbatim}
\usepackage{enumitem}
\usepackage{soul,physics}
\usepackage{subfigure}
\usepackage{dsfont} 

\usepackage{tikz}
\usetikzlibrary{arrows.meta, decorations.markings, decorations.pathmorphing, positioning}

\usepackage[driverfallback=dvipdfm]{hyperref}
\hypersetup{pdfpagemode=FullScreen,colorlinks=true,breaklinks,urlcolor=blue,linkcolor=blue,citecolor=blue}

\begin{document}
 
  \title{Enhancing Many-Body Chaos via Entropy Injection from Environment}

  \author{Yuke Zhang}
  \thanks{These authors contributed equally to this work.}
  \affiliation{Department of Physics \& State Key Laboratory of Surface Physics, Fudan University, Shanghai, 200438, China}

  \author{Wenbo Zhou}
  \thanks{These authors contributed equally to this work. \\ wb62025@outlook.com}
  \affiliation{Department of Physics \& State Key Laboratory of Surface Physics, Fudan University, Shanghai, 200438, China}

  \author{Pengfei Zhang}
  \thanks{PengfeiZhang.physics@gmail.com}
  \affiliation{Department of Physics \& State Key Laboratory of Surface Physics, Fudan University, Shanghai, 200438, China}
  \affiliation{Hefei National Laboratory, Hefei 230088, China}

  \date{\today}

  \begin{abstract}
   In closed quantum systems, local information spreads throughout the entire system and becomes highly complex under unitary evolution. In contrast, when the system is embedded in an environment, system-environment coupling can transfer information from the system into the environment, thereby reducing the rate of complexity growth within the system. This leads to the environment-induced scrambling transition established in previous works. In this work, we identify entropy injection from the environment as a different physical process that instead enhances many-body chaos. Our setup consists of coupling a system that is already in equilibrium with one environment to another environment, which serves as an entropy reservoir and drives the system into a non-equilibrium state. When entropy flows into the system through either heat transfer or particle transfer, the effective Hilbert space explored by the system enlarges, a mechanism that can enhance many-body chaos. We explicitly demonstrate this idea by constructing a solvable complex Brownian SYK model, in which both the relaxation toward the steady state and the steady-state quantum Lyapunov exponent can be computed analytically. Our results provide a controllable mechanism for tuning quantum scrambling through entropy flow in quantum many-body systems coupled to environments.
  \end{abstract}
  
  \maketitle

  \emph{ \color{blue}Introduction.--} Under unitary evolution, initially local operators spread over an increasing number of degrees of freedom in a quantum many-body system. This growth in operator size describes information scrambling~\cite{Hayden_2007,Sekino_2008,Shenker_2014,Kitaev:2014talk,shenker2015stringyeffectsscrambling,Maldacena_2016,PhysRevLett.117.111601,yoshida2017efficientdecodinghaydenpreskillprotocol,Nahum_2018,von_Keyserlingk_2018,Roberts_2018,Nahum_2018,von_Keyserlingk_2018,G_rttner_2017,Li_2017,Landsman_2019,Blok_2021,Jafferis:2022traversable,Li_2023}, the process by which local information becomes highly nonlocal and effectively inaccessible to local probes. It plays a central role in non-equilibrium quantum dynamics and has attracted broad interest across condensed matter physics, quantum information, and high-energy physics~\cite{Swingle2018,Qi2018,Chowdhury_2022}. In generic systems with sufficiently long-range interactions, information scrambling is characterized by the exponential growth of operator size at early times, a signature of quantum many-body chaos governed by the quantum Lyapunov exponent. This exponential amplification has enabled the identification of an effective theory in terms of collective modes known as scramblons~\cite{Kitaev_2018a,Gu_2019,Stanford:2021bhl,Gu:2021xaj,stanford2023scramblonloops,Zhang_2023a,PhysRevLett.132.060201,liu2026distinguishingcoherentincoherenterrors}, which has been experimentally demonstrated in solid-state nuclear magnetic resonance systems~\cite{cg3f-rggs}.

  Meanwhile, the development of quantum science and technology has inspired researchers to consider quantum systems that interact with external apparatuses, including open quantum systems and systems subject to repeated measurements. In particular, a series of studies has considered information scrambling in systems embedded in an environment~\cite{Chen_2017,Syzranov_2018,Zhang_2019o,Tuziemski_2019,almheiri2019universalconstraintsenergyflow,Bhattacharya_2022,schuster2022operatorgrowthopenquantum,Liu_2023,Bhattacharjee_2023,Bhattacharjee_2024,Zhang_2023,Weinstein_2023,Zhang:2024vsa,Zhou:2024osg,garcíagarcía2024lyapunovexponentsignaturedissipative,Zhang:2025ckq,Liu:2026mox}, where dissipation induced by the system-environment coupling competes with intrinsic chaotic dynamics. This leads to an environment-induced scrambling transition~\cite{Zhang_2023,Weinstein_2023,Zhang:2024vsa,Zhou:2024osg,garcíagarcía2024lyapunovexponentsignaturedissipative,Zhang:2025ckq,Liu:2026mox}, in which the quantum Lyapunov exponent remains finite only below a threshold of the system-environment coupling. Recently, Liu et al. further investigated the effect of introducing continuous monitoring into a system already coupled to an environment~\cite{Liu:2026mox}, modeling the monitoring dynamics using a Lindbladian master equation~\cite{10.1093/acprof:oso/9780199213900.001.0001,Manzano_2020}. Their numerical results suggest that monitoring can enhance or even induce quantum many-body chaos, pointing to a mechanism distinct from that of dissipation. Nevertheless, a clear theoretical understanding of this phenomenon remains elusive.

  In this work, we identify entropy injection as the underlying mechanism by which coupling to external systems enhances quantum many-body chaos, whether through coupling to an environment or through continuous monitoring. Our setup consists of coupling a system already embedded in an environment to a second environment with a different temperature or chemical potential, thereby driving the system into a non-equilibrium steady state. During the evolution, entropy injection enlarges the effective Hilbert-space dimension accessible to the dynamics, which in turn can enhance the quantum Lyapunov exponent. We demonstrate these predictions explicitly in a solvable complex Brownian Sachdev-Ye-Kitaev (SYK) model~\cite{Kitaev2015talk1,Sachdev_1993,PhysRevD.94.106002,saad2019semiclassicalrampsykgravity,S_nderhauf_2019,Liu_2021,Jian_2021,PhysRevLett.127.140601,PhysRevLett.134.156301,Davison_2017,Bulycheva_2017,Chaturvedi_2018,Gu_2020,Chen:2020bmq,Zhangp_2023}, where both the relaxation dynamics toward the non-equilibrium steady state and the corresponding quantum Lyapunov exponent can be computed analytically. Our results provide a new route toward the precise control of information scrambling in quantum many-body systems.

  \begin{figure}[t]
    \centering
    \includegraphics[width=0.87\linewidth]{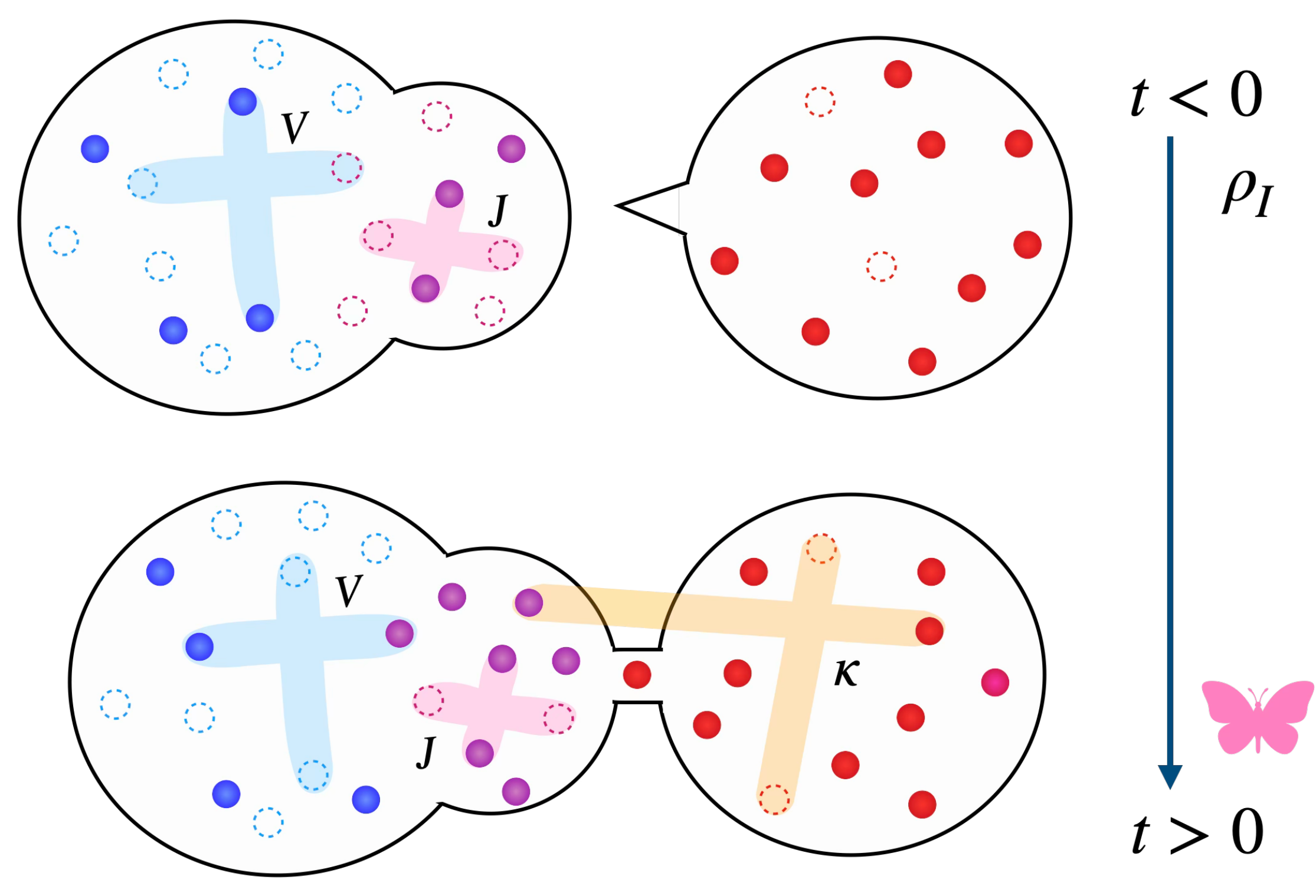}
    \caption{Schematic illustration of the setup. Filled and empty circles represent occupied and unoccupied modes, respectively, with $J$ denoting the intra-system coupling strength. For $t<0$, the system is coupled to a single bath with strength $V$. At $t=0$, a second bath is activated with coupling strength $\kappa$, inducing a change in the system's density accompanied by entropy injection. At later times ($t>0$), the system relaxes into a steady state, where the quantum Lyapunov exponent is evaluated as a signature of the quantum butterfly effect.}
    \label{fig:num1}
  \end{figure}
  \emph{ \color{blue}Setup.--} We first introduce the setup considered in this work. We consider a system consisting of $N$ complex fermions with annihilation operators $c_j$, where $j\in\{1,2,\cdots,N\}$. They satisfy the canonical anticommutation relation $\{c_j,c_k^\dagger\}=\delta_{jk}$. Nevertheless, the physical mechanism discussed in this work naturally generalizes to bosonic or spin systems. The system $S$ is in equilibrium with an environment $E_1$ described by $M$ fermionic modes $\xi_p$ with $M\gg N$, where $p\in\{1,2,\cdots,M\}$. To be concrete, we use a solvable Brownian SYK model with charge conservation~\cite{saad2019semiclassicalrampsykgravity,S_nderhauf_2019,Liu_2021,Jian_2021,PhysRevLett.127.140601,PhysRevLett.134.156301,Davison_2017,Bulycheva_2017,Chaturvedi_2018,Gu_2020,Chen:2020bmq,Zhangp_2023,Zhang_2023,Zhang:2024vsa} as an explicit example. The Hamiltonian is given by
  \begin{equation}\label{eq:H}
  \begin{aligned}
  H= &\sum_{i<j,k<l}J_{ijkl}(t)c^\dagger_ic^\dagger_jc_kc_l\\&+\sum_{ip,q<o}\Big(v_{ipqo}(t)c^\dagger_i\xi^\dagger_p\xi_q\xi_o+\text{h.c.}\Big).
  \end{aligned}
  \end{equation}
  Here, the random couplings $J_{ijkl}(t)$ and $v_{ipqo}(t)$ with different indices are independent Brownian variables with zero mean and variances given by
  \begin{equation}
  \begin{aligned}
  \overline{J_{ijkl}(t_1)J_{ijkl}(t_2)^*}&=2J\delta(t_{12})/N^3,\\
  \overline{v_{ipqo}(t_1)v_{ipqo}(t_2)^*}&={2V}\delta(t_{12})/{M^3}.
  \end{aligned}
  \end{equation}
  We focus on the thermodynamical limit $N\rightarrow \infty$, where the properties of the system can be analyzed using the large-$N$ expansion. We can further introduce arbitrary intrinsic dynamics of the bath, as this does not affect the subsequent discussions. The model preserves the total charge, while charge can still be exchanged between the system and the environment. Therefore, when the total system reaches thermal equilibrium, the Gibbs ensemble is parameterized by a single chemical potential, $\rho=e^{\mu Q}/Z$, with total particle number $Q=\sum_i c_i^\dagger c_i+\sum_p\xi_p^\dagger\xi_p$ and $Z=\mathrm{tr}[e^{\mu Q}]$. It is then straightforward to determine the particle number as $\langle \xi_p^\dagger \xi_p\rangle=\langle c_i^\dagger c_i\rangle=\frac{1}{e^{-\mu}+1}\equiv n$. The corresponding entropy per site in the system $S$ is given by $\mathcal{S}/N =-n\ln n-(1-n)\ln(1-n)$, which characterizes the effective dimension $e^{\mathcal{S}}$ of the constrained Hilbert space. It reaches its maximum at $n = 1/2$ and vanishes as $n \rightarrow 0$ or $n \rightarrow 1$.

  In Ref.~\cite{Zhang_2023,Zhang:2024vsa}, it has been established that a system in thermal equilibrium exhibits a scrambling transition as the system-environment coupling $V$ is increased. For $V<2J$, the dynamics are dominated by intrinsic chaotic behavior, leading to a positive quantum Lyapunov exponent $\varkappa_0 = 2\gamma_0-\gamma_1$. Here, we have introduced $\gamma_0=J n(1-n)$ and $\gamma_1 = Vn(1-n)$. The factor $n(1-n)$ describes the suppression of quantum chaos due to the reduced dimension of the accessible Hilbert space at low and high densities~\cite{Chen:2020bmq,Zhangp_2023,Zhang:2024vsa}. In contrast, when $V>2J$, coupling to the environment rapidly transfers information into $E_1$, and operator growth within the system $S$ ceases. Signatures of the scrambling transition are also reflected in the entropy dynamics~\cite{Zhang:2025ckq} and in the system’s ability to process quantum information~\cite{Zhou:2024osg}.

  Next, we couple the system to a second environment $E_2$, described by fermionic modes $\eta_p$ with $p \in \{1,2,\cdots,M\}$. For now, we assume that the coupling to the second environment takes a similar form as in \eqref{eq:H}:
   \begin{equation}\label{eq:H2}
  \begin{aligned}
  H_{SE_2}= \Theta(t)\sum_{ip,q<o}\Big(\kappa_{ipqo}(t)c^\dagger_i\eta^\dagger_p\eta_q\eta_o+\text{h.c.}\Big).
  \end{aligned}
  \end{equation}
  The corresponding coupling strength, denoted by $\kappa_{ipqm}(t)$, has a variance
  \begin{equation}
  \begin{aligned}
  \overline{\kappa_{ipqo}(t_1)\kappa_{ipqo}(t_2)^*}&={2\kappa}\delta(t_{12})/{M^3}.
  \end{aligned}
  \end{equation}
  Here, $\Theta(t)$ is the Heaviside step function, indicating that the coupling to $E_2$ is turned on only for $t > 0$. The initial state of $E_2$ is taken to be an equilibrium state with an independent chemical potential $\nu$, which is generally different from $\mu$. This leads to a particle number $\langle \eta_p^\dagger \eta_p\rangle=\frac{1}{e^{-\nu}+1}\equiv m$. As a result, the system $S$ evolves under non-equilibrium dynamics. For any finite evolution time $t$, the total system does not reach thermal equilibrium, since the back-action of the system on the environments is negligible. Mathematically, this is guaranteed by a vanishing self-energy for both $\xi_p$ and $\eta_p$ in the limit $M \gg N$. Nevertheless, we expect the system to reach a non-equilibrium steady state. Our main focus is to analyze the relaxation toward this steady state and the resulting quantum Lyapunov exponent.

  \emph{ \color{blue}Relaxation dynamics.--} We now investigate how the system reaches the non-equilibrium steady state. We introduce the standard real-time Green’s functions as $G^>(t_1,t_2)=-i \langle c_i({t_1})c_i^\dagger (t_2)\rangle$ and $G^<(t_1,t_2)=i \langle c_i^\dagger (t_2)c_i({t_1})\rangle$ for the system $S$, and analogously $G^{\gtrless}_{1/2}$ for the environments $E_1$ and $E_2$. The evolution of $G^>$ and $G^<$ is governed by the Kadanoff-Baym equation~\cite{Eberlein_2017,Zhang_2019,Kamenev_2023,Stefanucci_vanLeeuwen_2025}:
  \begin{equation}
  \begin{aligned}
  i\partial_{t_1}G^{\gtrless}&=\Sigma^R\circ G^{\gtrless}+\Sigma^{\gtrless}\circ G^A,\\
  -i\partial_{t_2} G^{\gtrless}&=G^R\circ \Sigma^{\gtrless}+G^{\gtrless}\circ \Sigma^A.
  \end{aligned}
  \end{equation}
  Here, $\circ$ denotes the convolution in the time domain. $\Sigma^{\gtrless}$ are the real-time self-energies, which receive contributions from the standard melonic diagrams in SYK-like models ($t_1,t_2>0$)~\cite{PhysRevD.94.106002}:
  \begin{equation}
  \begin{aligned}
  \Sigma^{\gtrless}&=\delta(t_{12}) \Big[J(G^{\gtrless})^2 G^{\lessgtr}+ V (G^{\gtrless}_1)^2 G^{\lessgtr}_1+ \kappa (G^{\gtrless}_2)^2 G^{\lessgtr}_2\Big].
  \end{aligned}
  \end{equation}
  Here, we omit the arguments of the Green’s functions for conciseness. The equal-time Green’s functions of the environments are determined by the particle number, for example, $G_1^{>} = -i(1-n)$ and $G_1^{<} = in$. We have also introduced the retarded and advanced Green’s functions $G^{R/A} = \pm \Theta(\pm t_{12})(G^> - G^<)$, and a similar relation holds for the self-energy $\Sigma^{R/A}$.

  We focus on the diagonal components $G^>(t,t) = -i(1-f(t))$ and $G^<(t,t) = if(t)$, whose relation is guaranteed by the canonical commutation relations. For Brownian models, their evolution effectively decouples from the off-diagonal components, leading to an exact quantum Boltzmann equation:
  \begin{equation}
  \frac{df(t)}{dt}=\gamma_1(n-f(t))+\gamma_2 (m-f(t)).
  \end{equation}
  The detailed derivation is presented in the Supplementary Material~\cite{SM}. Here, we have introduced $\gamma_2 = \kappa m(1-m)$. The two terms on the right-hand side describe the charge transfer induced by the coupling to $E_1$ and $E_2$, respectively. The equation is solved under the initial condition $f(0) = n$, which leads to
  \begin{equation}
    f(t)=\frac{\gamma_1n+\gamma_2m}{\gamma_1+\gamma_2}+\frac{\gamma_2(n-m)}{\gamma_1+\gamma_2}e^{-(\gamma_1+\gamma_2)t}.
  \end{equation}
  In particular, we find the steady-state particle number to be $f(\infty)=\frac{\gamma_1n+\gamma_2m}{\gamma_1+\gamma_2}\equiv f$. We highlight three distinct regimes as $\kappa$ increases, assuming $n < 1/2$: (1) When $m<n<1/2$, the density of system $S$ decreases as $\kappa$ increases, and the accessible Hilbert-space dimension $e^{\mathcal{S}}$ decreases monotonically. (2) When $n<m<1/2$, the density increases together with the accessible Hilbert-space dimension, indicating entropy injection from the environment. (3) When $n<1/2<m$, the density increases, while the accessible Hilbert-space dimension exhibits nonmonotonic behavior. In both cases (2) and (3), a small $\kappa$ enhances the accessible Hilbert-space dimension and therefore may promote information scrambling, as explicitly demonstrated in later discussions. In contrast, for $n = 1/2$, the entropy and accessible Hilbert-space dimension both decrease for any $m$.
  
  \begin{figure}[t]
    \centering
    \includegraphics[width=0.98\linewidth]{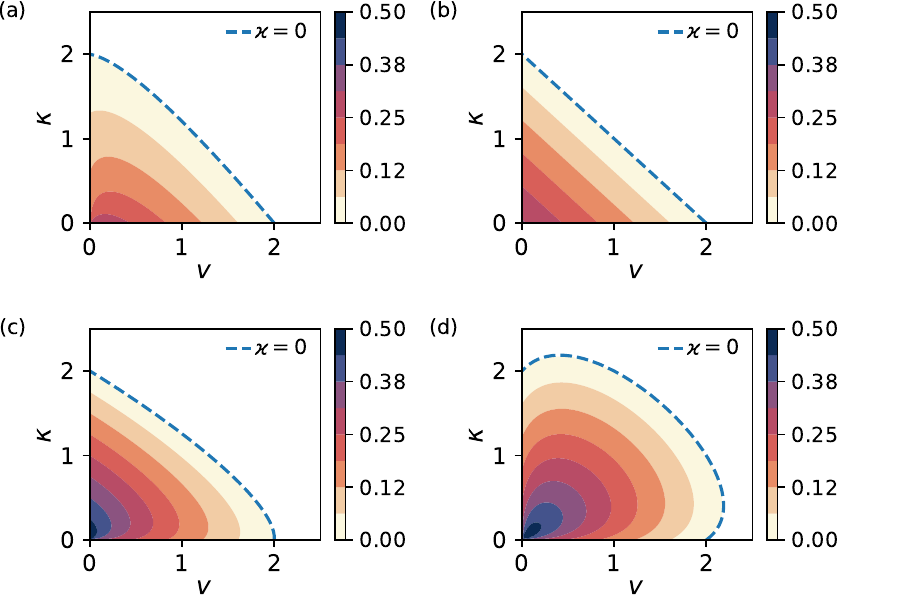}
    \caption{Contour plots of the quantum Lyapunov exponent for the complex Brownian SYK model with the original $SE_2$ coupling \eqref{eq:H2}, for (a) $m = 0.1$, (b) $m = 0.2$, (c) $m = 0.48$, and (d) $m = 0.8$. We fix $n = 0.2$ and $J = 1$ in all panels. The dashed lines denote the phase boundary between the scrambling phase and the dissipative phase.}
    \label{fig:num1}
  \end{figure}

  \emph{ \color{blue}Many-body chaos.--} We next calculate the quantum Lyapunov exponent in the nonequilibrium steady state, with particular focus on whether the coupling to $E_2$ enhances many-body chaos. We define the retarded out-of-time-ordered correlator (OTOC) with respect to the steady-state density matrix $\rho_{\mathrm{s.s.}}$ as~\cite{Liu:2026mox}
  \begin{equation}
  \begin{aligned}
  F_0(t)&=\text{tr}[\rho_{\mathrm{s.s.}}\{c_j^\dagger(0),c_k(t)\}\{c_j(0),c_k^\dagger(t)\}],\\
  F_1(t)&=-\text{tr}[\rho_{\mathrm{s.s.}}\{c_j^\dagger(0),c_k^\dagger(t)\}\{c_j(0),c_k(t)\}].
  \end{aligned}
  \end{equation}
  Following the standard methodology, we express the OTOCs using doubled Keldysh contour and view the operators inserted at $t=0$ as perturbations~\cite{Aleiner:2016eni,Gu:2021xaj}. The operators at time $t$ then becomes a two-point function on the perturbed background. As a consequence, the OTOCs can be calculated by investigating the Schwinger-Dyson equation on the doubled Keldysh contour. Leaving the details to the Supplementary Material~\cite{SM}, the equation satisfied by the OTOCs before the scrambling time reads
  \begin{equation}
  \frac{d\bm{F}(t)}{dt} =(\tilde{\gamma}_0-\gamma_1-\gamma_2)\bm{F}(t)+\tilde\gamma_0\sigma_x \bm{F}(t).
  \end{equation}
  Here, we introduce the two-component OTOC vector $\bm{F}(t)=(F_0(t)/(1-f),-F_1(t)/f)^T$ and define $\tilde{\gamma}_0 = J f(1-f)$. The quantum Lyapunov exponent $\varkappa$ is identified with the largest eigenvalue of the right-hand side, yielding
  \begin{equation}\label{eq:res}
  \varkappa=2\tilde{\gamma}_0-\gamma_1-\gamma_2=\varkappa_0+2(\tilde{\gamma}_0-\gamma_0)-\gamma_2.
  \end{equation}
  The result \eqref{eq:res} naturally separates two distinct contributions to the difference between $\varkappa$ and $\varkappa_0$. First, the renormalization of the intrinsic growth rate, encoded in the shift $\tilde{\gamma}_0 - \gamma_0 = J\big(f(1-f) - n(1-n)\big)$, reflects a modification of the particle density, or equivalently, a change in the accessible Hilbert-space dimension. When the accessible Hilbert-space dimension increases, corresponding to entropy injected from the environment, this term can enhance quantum many-body chaos. Second, the term $-\gamma_2$ captures the direct leakage of information into the additional environment, which always suppresses the growth of operator size and thus reduces the quantum Lyapunov exponent. 

  We can determine the conditions under which the Lyapunov exponent increases upon introducing a small $\kappa$ by performing a perturbative expansion:
  \begin{equation}
  \varkappa=\varkappa_0+\left(\frac{2 (1-2 n) (m-n)}{(1-n) n}-1\right)\gamma_2+ O(\kappa^2).
  \end{equation}
  Therefore, the deviation between $m$ and $n$ must be sufficiently large to observe an enhancement at small $\kappa$. In contrast, for large $\kappa$, the particle number saturates to $f = m$, and dissipation always becomes the dominant effect. As a result, the quantum Lyapunov exponent decreases with increasing $\kappa$ in the large-$\kappa$ regime. We further illustrate these results using contour plots of the quantum Lyapunov exponent in FIG.~\ref{fig:num1}. The shape of the phase boundary clearly shows that coupling to $E_2$ can enhance or even induce many-body chaos at small $\kappa$, while large $\kappa$ always suppresses chaos, consistent with our analysis. A similar enhancement can also occur at small $V$, where the roles of the two environments are interchanged. In the Supplementary Material~\cite{SM}, we further analyze the late-time behavior of the OTOCs using scramblon theory. The results similarly reveal a non-monotonic dependence of the OTOC saturation value on the coupling $\kappa$.

  The mechanism identified in the solvable SYK model admits a natural generalization to broader classes of quantum many-body systems. We consider systems with conserved quantities such as particle number or energy, which are exchanged with an initial environment $E_1$ to reach thermal equilibrium at fixed chemical potential or temperature. We then couple the system to a second environment $E_2$ with different thermodynamic parameters, generating a thermodynamic force associated with gradients in chemical potential or temperature. This drive induces an entropy flux into the system when the transfer of conserved quantities increases the system’s local entropy. It effectively enlarges the set of dynamically accessible many-body configurations consistent with the conservation laws, thereby increasing the phase space available for operator growth. Scrambling dynamics is thus governed not only by information loss to the environment, but also by the competition between dissipation and entropy injection. This suggests a general principle: entropy flow into a quantum many-body system can serve as a resource that enhances scrambling, complementing the conventional environment-induced suppression of chaos.
  
  \begin{figure}[t]
    \centering
    \includegraphics[width=0.98\linewidth]{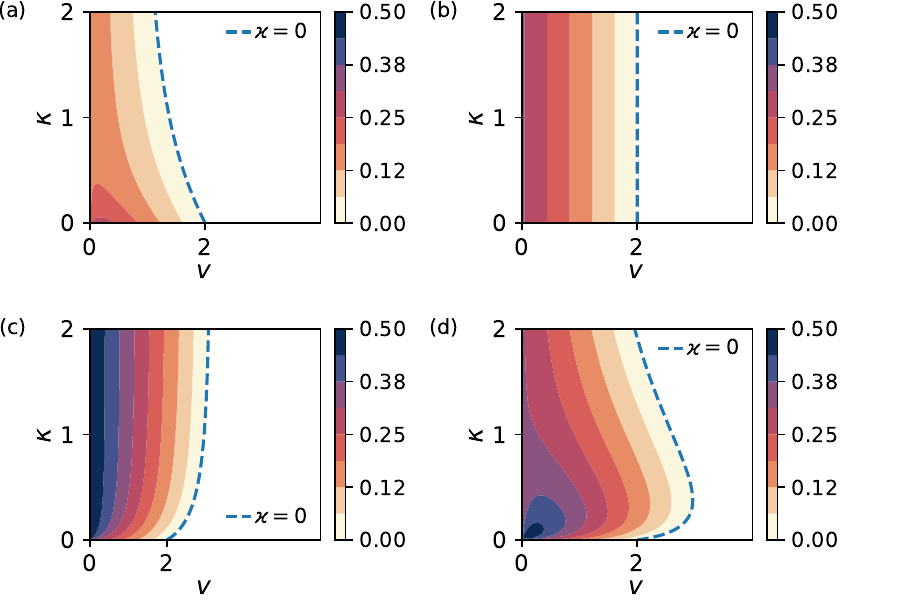}
    \caption{Contour plots of the quantum Lyapunov exponent for the complex Brownian SYK model with the pair-hopping coupling \eqref{eq:H22}, for (a) $m = 0.1$, (b) $m = 0.2$, (c) $m = 0.48$, and (d) $m = 0.8$. We fix $n = 0.2$ and $J = 1$ in all panels. The dashed lines denote the phase boundary between the scrambling phase and the dissipative phase.}
    \label{fig:num2}
  \end{figure}

  \emph{ \color{blue}Coupling without dissipation.--} With this understanding, we can further design alternative system-environment couplings that suppress the dissipation induced by the environment while maintaining the enhancement of many-body chaos even at large $\kappa$. The modified coupling term is a pair-hopping between $S$ and $E_2$, which reads
  \begin{equation}\label{eq:H22}
  \begin{aligned}
  H_{SE_2}= \Theta(t)\sum_{i<j,p<q}\Big(\kappa_{ijqq}(t)c^\dagger_ic^\dagger_j\eta_p\eta_q+\text{h.c.}\Big).
  \end{aligned}
  \end{equation}
  The variance of the coupling strength $\kappa_{ijqq}(t)$ reads
  \begin{equation}
  \begin{aligned}
  \overline{\kappa_{ijqq}(t_1)\kappa_{ijqq}(t_2)}&={2\kappa}\delta(t_{12})/{M^3}.
  \end{aligned}
  \end{equation}
  The coupling \eqref{eq:H22} preserves the operator size within the system: taking the commutator $[c_k, H_{SE_2}] \sim \sum \kappa_{kjpq} c^\dagger_j \eta_p \eta_q$, we see that it still contains a system operator $c^\dagger_j$. Therefore, this system-environment coupling does not lead to a loss of operator size and hence does not induce dissipation. Following the same strategy as in the previous sections, we are able to calculate the quantum Lyapunov exponent analytically, as presented in the Supplementary Material~\cite{SM}. Here, we only provide the numerical results in Fig.~\ref{fig:num2}. The results exhibit three distinct regimes, determined by both $m-n$ and $m-\tfrac{1}{2}$, as discussed in the previous sections. This clearly demonstrates the ability to control many-body chaos without introducing dissipation, highlighting potential applications in quantum information processing and quantum devices.

  \emph{\color{blue}Discussions.--} In this work, we show that coupling to external environments can enhance, rather than suppress, quantum many-body chaos through entropy injection. By driving a system into a nonequilibrium steady state using environments with different chemical potentials, entropy influx increases the accessible many-body state space and promotes operator growth. Using a solvable complex Brownian SYK model, we analytically characterized both the relaxation dynamics and the quantum Lyapunov exponent, revealing the competition between entropy injection and dissipation. We further showed that appropriately designed couplings can suppress dissipative effects while retaining the enhancement of scrambling. Our results establish entropy flow as a controllable resource for manipulating quantum chaos. 

  We conclude our work with a few remarks on future directions. First, it has been established that information scrambling can enable novel protocols in quantum many-body systems, such as wormhole teleportation~\cite{Gao_2017,Maldacena_2017}. It would therefore be interesting to investigate whether the mechanism identified here can also enhance the fidelity of teleportation signals through entropy injection. Second, it would be interesting to explore whether entropy injection can modify other dynamical diagnostics associated with quantum chaos, such as Krylov complexity~\cite{Parker_2019,Nandy_2025,rabinovici2025krylovcomplexity}. Third, it would be interesting to explore the interplay between entropy injection and measurement-induced dynamics~\cite{Li_2019,Szyniszewski_2019,Skinner_2019}, where monitoring and feedback may provide additional routes for controlling operator growth and many-body chaos in open quantum systems. We leave these directions for future work.

  \textit{Acknowledgement.}
  This project is supported by the Shanghai Rising-Star Program under grant number 24QA2700300, the NSFC under grant 12374477, the Quantum Science and Technology-National Science and Technology Major Project 2024ZD0300101, and the Xuemin Institute of Advanced Studies, Fudan University.

\bibliography{ref.bib}




\clearpage
\onecolumngrid

\vspace{12pt} 
\begin{center}
\bfseries\large Supplementary Material for:\\ Enhancing Many-Body Chaos via Entropy Injection from Environment
\end{center}
\vspace{6pt} 

\setcounter{equation}{0}
\setcounter{figure}{0}
\setcounter{table}{0}
\renewcommand{\theequation}{S\arabic{equation}}
\renewcommand{\thefigure}{S\arabic{figure}}
\renewcommand{\thetable}{S\arabic{table}}

This Supplementary Material consists of two parts. In Sec.~A, we provide a detailed derivation of the relaxation dynamics for three SYK-like models. In Sec.~B, we further investigate the behavior of out-of-time-ordered correlators (OTOCs) in these models, focusing on the quantum Lyapunov exponent and the second freeness distance.

\section{A. Relaxation dynamics}\label{Relax}
In this section, we calculate the relaxation dynamics from the initial state to the non-equilibrium steady state. Specifically, we provide a detailed derivation of the quantum Boltzmann equation (7) in the main text, and calculate the Green's functions in the non-equilibrium steady state. We mainly focus on three SYK-like models. Model 1, introduced as the baseline model in the main text, features a four-body system interaction and is hereafter referred to as CBSYK(4). Model 2 shares the same system interaction form as Model 1, but is designed to be free of dissipation. In contrast, the interaction term in Model 3 comprises three system operators and one bath operator, which we designate as CBSYK(3,1).

\subsection{Model 1: CBSYK(4) with dissipation coupling}\label{Beq1}
We introduce the standard real-time Green's functions on the Schwinger-Keldysh contour as follows:
\begin{equation}\label{eq:Greensfun}
G(t_1,t_2) = 
\begin{pmatrix}
G^T(t_1,t_2) & G^<(t_1,t_2)  \\
G^>(t_1,t_2) & G^{\Tilde{T}}(t_1,t_2) \\
\end{pmatrix}
=
\begin{pmatrix}
-i\langle\psi^1_i(t_1)\overline{\psi}^1_i(t_2)\rangle & -i\langle\psi^1_i(t_1)\overline{\psi}^2_i(t_2)\rangle  \\
-i\langle\psi^2_i(t_1)\overline{\psi}^1_i(t_2)\rangle & -i\langle\psi^2_i(t_1)\overline{\psi}^2_i(t_2)\rangle \\
\end{pmatrix},
\end{equation}
where the indices $1$ and $2$ denote the forward and backward evolution branches, respectively. The corresponding self-energy matrix is given by
\begin{equation}\label{eq:Selfenergy}
\Sigma(t_1,t_2) = 
\begin{pmatrix}
\Sigma^T(t_1,t_2) & -\Sigma^<(t_1,t_2)  \\
-\Sigma^>(t_1,t_2) & \Sigma^{\Tilde{T}}(t_1,t_2) \\
\end{pmatrix}.
\end{equation}
In this matrix representation, the Schwinger-Dyson equations can be written as
\begin{equation}\label{eq:Schwinger-Dyson1}
\begin{pmatrix}
i\partial_t-\Sigma^T & \Sigma^<  \\
\Sigma^> & -i\partial_t-\Sigma^{\Tilde{T}} \\
\end{pmatrix}\circ
\begin{pmatrix}
G^T & G^<  \\
G^> & G^{\Tilde{T}} \\
\end{pmatrix}=\mathds{1},
\end{equation}
and
\begin{equation}\label{eq:Schwinger-Dyson2}
\begin{pmatrix}
G^T & G^<  \\
G^> & G^{\Tilde{T}} \\
\end{pmatrix}\circ
\begin{pmatrix}
i\partial_t-\Sigma^T & \Sigma^<  \\
\Sigma^> & -i\partial_t-\Sigma^{\Tilde{T}} \\
\end{pmatrix}
=\mathds{1}.
\end{equation}
Here, the explicit time arguments of the Green's functions and self-energies are omitted for conciseness. We have also utilized the fact that the inverse bare Green's function for the SYK model takes the form $G_0^{-1} = (i\partial_t, -i\partial_t)$. By combining Eqs.~(\ref{eq:Schwinger-Dyson1}) and (\ref{eq:Schwinger-Dyson2}), we arrive at the Kadanoff-Baym equations:
  \begin{equation}
  \begin{aligned}
  i\partial_{t_1}G^{\gtrless}&=\Sigma^R\circ G^{\gtrless}+\Sigma^{\gtrless}\circ G^A,\\
  -i\partial_{t_2} G^{\gtrless}&=G^R\circ \Sigma^{\gtrless}+G^{\gtrless}\circ \Sigma^A.
  \end{aligned}
  \end{equation}
Here, we have introduced the retarded and advanced Green's functions, defined as $G^{R/A}(t_1,t_2) = \pm \Theta(\pm t_{12})(G^> - G^<)$, along with analogous relations for the self-energies $\Sigma^{R/A}$. 

\begin{figure}[htbp]
    \centering
    \includegraphics[width=0.85\linewidth]{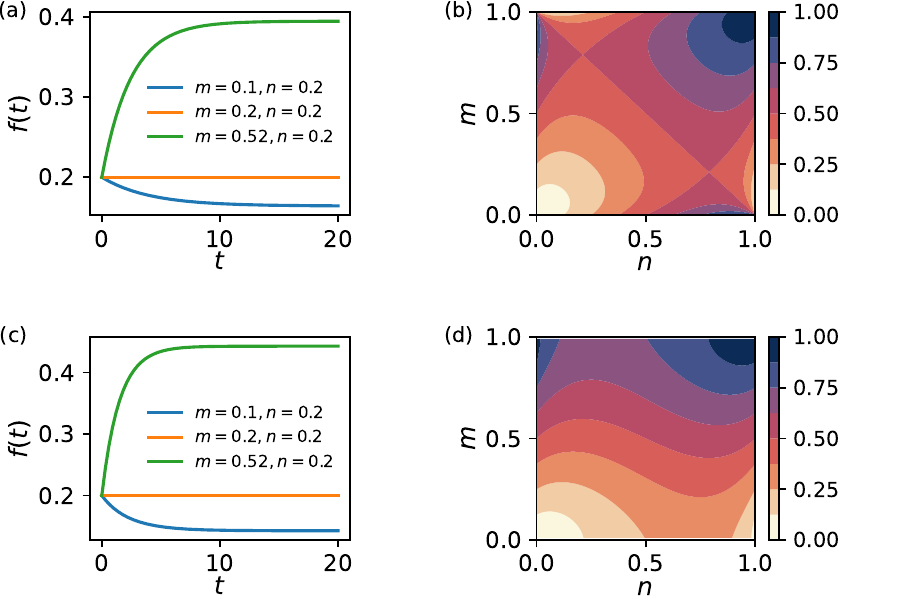}
    \caption{Evolution of the density $f(t)$ for (a) Model 1, and (c) Models 2 and 3. Panels (b) and (d) display the contour plots of the steady-state system density corresponding to the models in (a) and (c), respectively. Here, the parameters are fixed at $J=1$, $v=1$, and $\kappa=1$. The colorbar represents the density $f$ with a resolution of approximately $0.06$.}
    \label{fig:curve}
  \end{figure}

Next, we rewrite these equations in terms of the center-of-mass time $t = \frac{t_1+t_2}{2}$ and the relative time $s = t_1 - t_2$, which yields
\begin{equation}\label{eq:BoltzT}
\begin{aligned}
    i\partial_tG^{\gtrless}(t,s)=&\tilde\Sigma^R(t+s/2)G^{\gtrless}(t,s)+\tilde\Sigma^{\gtrless}(t+s/2)G^A(t,s)\\
    &-G^{R}(t,s)\tilde\Sigma^{\gtrless}(t-s/2)-G^{\gtrless}(t,s)\tilde\Sigma^A(t-s/2).
\end{aligned}
\end{equation}
\begin{equation}\label{eq:Boltzt}
\begin{aligned}
    i\partial_sG^{\gtrless}(t,s)=&\frac{1}{2}\tilde\Sigma^{R}(t+s/2)G^{\gtrless}(t,s)+\frac{1}{2}\tilde\Sigma^{\gtrless}(t+s/2)G^A(t,s)\\
    &+\frac{1}{2}G^{R}(t,s)\tilde\Sigma^{\gtrless}(t-s/2)+\frac{1}{2}G^{\gtrless}(t,s)\tilde\Sigma^A(t-s/2).
\end{aligned}
\end{equation}
Here, $\Sigma^{\gtrless}$ are the real-time self-energies, which receive contributions from the standard melonic diagrams in SYK-like models ($t_1,t_2>0$):
  \begin{equation}
  \begin{aligned}
  \Sigma^{\gtrless}&=\delta(t_{12}) \Big[J(G^{\gtrless})^2 G^{\lessgtr}+ V (G^{\gtrless}_1)^2 G^{\lessgtr}_1+ \kappa (G^{\gtrless}_2)^2 G^{\lessgtr}_2\Big]\equiv\delta(t_{12})\tilde\Sigma^{\gtrless}.
  \end{aligned}
  \end{equation}
We first consider Eq.~(\ref{eq:BoltzT}), which governs the relaxation dynamics. Upon setting $s = 0$, we find that the evolution of the diagonal components effectively decouples from the off-diagonal ones, which is a hallmark of the Brownian SYK model. We further express these diagonal components in terms of the distribution function $f(t)$ : $G^>(t,t) = -i(1 - f(t))$ and $G^<(t,t) = if(t)$. This yields the exact quantum Boltzmann equation:
\begin{equation}\label{eq:Boltz1}
\frac{df(t)}{dt}=\gamma_1(n-f(t))+\gamma_2 (m-f(t)).
\end{equation}
With the initial condition $f(0) = n$, the analytical solution is given by
\begin{equation}
    f(t)=\frac{\gamma_1n+\gamma_2m}{\gamma_1+\gamma_2}+\frac{\gamma_2(n-m)}{\gamma_1+\gamma_2}e^{-(\gamma_1+\gamma_2)t}.
\end{equation}
Consequently, the steady-state particle number is obtained as $f(\infty) = \frac{\gamma_1 n + \gamma_2 m}{\gamma_1 + \gamma_2} \equiv f$. We present the full relaxation dynamics alongside the steady-state particle density in~FIG.\ref{fig:curve}(a,b).

Next, we turn our attention to the steady-state Green's functions, e.g., $G^<_{ss}(s) = -i \langle c_i(s) c_i^\dagger(0) \rangle_{ss}$. Eq.~(\ref{eq:Boltzt}) can be simplified as:
\begin{equation}\label{eq:simBoltzt}
\begin{aligned}
    i\partial_sG^{<}_{ss}(s)=&\frac{1}{2}\tilde\Sigma^{R}_{ss}G^{\gtrless}_{ss}(s)+\frac{1}{2}\tilde\Sigma_{ss}^{\gtrless}G^A_{ss}(s)
    +\frac{1}{2}G_{ss}^{R}(s)\tilde\Sigma_{ss}^{\gtrless}+\frac{1}{2}G_{ss}^{\gtrless}(s)\tilde\Sigma_{ss}^A.
\end{aligned}
\end{equation}
Here, $\tilde\Sigma^{\gtrless}_{ss}$ are calculated directly using the steady-state solution of Eq.~(\ref{eq:Boltz1}). Finally, we obtain $G^{R/A}_{ss}(s) = \mp i\Theta(\pm s)e^{-\Gamma s/2}$, $G_{ss}^<(s) = -fe^{-\Gamma |s|/2}$, and $G_{ss}^>(s) = (1 - f)e^{-\Gamma |s|/2}$. Here, we have introduced the total decay rate $\Gamma = \tilde\gamma_0 + \gamma_1 + \gamma_2$ for the quasi-particles.

\subsection{Model 2: CBSYK(4) without dissipation coupling}\label{Beq2}
We next consider an alternative system-environment coupling that suppresses the dissipation induced by the environment (see Eq.~(13) in the main text). The only difference from the previous model lies in the concrete form of the self-energies, which now read:
 \begin{equation}
 \begin{aligned}
 \Sigma^{\gtrless}&=\delta(t_{12}) \Big[J(G^{\gtrless})^2 G^{\lessgtr}+ V (G^{\gtrless}_1)^2 G^{\lessgtr}_1+ \kappa (G^{\gtrless}_2)^2 G^{\lessgtr}\Big].
 \end{aligned}
 \end{equation}
Following a similar procedure to that in the previous section, we obtain the relaxation dynamics equation:
\begin{equation}\label{eq:Boltz2}
\frac{\text{d}f(t)}{\text{d}t} =\gamma_1 \left[ n - f(t) \right] +\gamma_2\frac{ \left[ f(t)^2 - 2m f(t)^2 + (2f(t) - 1)m^2 \right]}{m(m - 1)}.
\end{equation}
Subject to the initial condition $f(0) = n$, the analytical solution is found to be
\begin{equation}
    f(t) = \frac{m \left[ \gamma_1(1-m) + 2\gamma_2m \right] - \sqrt{m(1-m)Q} \cdot \frac{\tanh(\omega t) + R}{1 + R \tanh(\omega t)}}{2\gamma_2(2m-1)},
\end{equation}
where $Q = m(1 - m)(\gamma_1^2 + 4\gamma_2^2) + 4\gamma_1\gamma_2(m^2 + n - 2mn)$, $R = \frac{m[\gamma_1(1-m) + 2\gamma_2 m] - 2\gamma_2 n(2m-1)}{\sqrt{m(1-m)Q}}$, and $\omega = \frac{\sqrt{Q}}{2\sqrt{m(1-m)}}$. The corresponding steady-state particle number is given by $f(\infty) = \frac{m[\gamma_1(1-m) + 2\gamma_2 m] - \sqrt{m(1-m)}Q}{2\gamma_2(2m-1)} \equiv f$. 

Furthermore, the steady-state Green's functions are given by $G_{ss}^{R/A}(s) = \mp i\Theta(\pm s)e^{-\Gamma s/2}$, $G_{ss}^<(s) = -fe^{-\Gamma |s|/2}$, and $G_{ss}^>(s) = (1 - f)e^{-\Gamma |s|/2}$, where the total quasi-particle decay rate is defined as $\Gamma = \tilde\gamma_0 + \gamma_1 + \gamma_2 \frac{f - 2fm + m^2}{m(1-m)}$. The full relaxation dynamics and the steady-state particle density for this model are presented in Fig.~1(c) and 1(d).
\subsection{Model 3: CBSYK(3,1) with dissipation coupling}\label{Beq3}
In this subsection, we further investigate a deformed model in which the full dynamics of OTOCs can be solved exactly, as will be discussed in detail in the next section. The corresponding Hamiltonian is given by
  \begin{equation}\label{eq:H31}
  H= \sum_{i<j,k,p}J_{ijkp}(t)\Big(c^\dagger_ic^\dagger_jc_k\xi_p+\text{h.c.}\Big)+\sum_{ip,q<o}\Big(v_{ipqo}(t)c^\dagger_i\xi^\dagger_p\xi_q\xi_o+\text{h.c.}\Big).
  \end{equation}
 \begin{equation}\label{eq:H312}
  \begin{aligned}
  H_{SE_2}= \Theta(t)\sum_{ip,q<o}\Big(\kappa_{ipqo}(t)c^\dagger_i\eta^\dagger_p\eta_q\eta_o+\text{h.c.}\Big).
  \end{aligned}
  \end{equation} 
  Here, the random couplings $J_{ijkl}(t)$ and $v_{ipqo}(t)$ with different indices are independent Brownian variables with zero mean and variances given by
  \begin{equation}
  \overline{J_{ijkl}(t_1)J_{ijkl}(t_2)^*}=J\delta(t_{12})/2N^2M,~~
  \overline{v_{ipqo}(t_1)v_{ipqo}(t_2)^*}={2V}\delta(t_{12})/{M^3},~~  \overline{\kappa_{ipqo}(t_1)\kappa_{ipqo}(t_2)^*}={2\kappa}\delta(t_{12})/{M^3}
  \end{equation}
The only difference from the baseline model lies in the explicit form of the self-energies, which are expressed as
\begin{equation}
\begin{aligned}
\Sigma^{\gtrless}&=\delta(t_{12}) \Big[J\left(G^{\gtrless} G^{\lessgtr}G_1^{\gtrless}+(G^{\gtrless})^2G_1^{\lessgtr}\right)+ V (G^{\gtrless}_1)^2 G^{\lessgtr}_1+ \kappa (G^{\gtrless}_2)^2 G^{\lessgtr}\Big].
\end{aligned}
\end{equation}
Following the same procedure as detailed above, we obtain the relaxation dynamics equation:
\begin{equation}\label{eq:Boltz3}
\frac{df(t)}{dt}=\gamma_1(n-f(t))+\gamma_2 (m-f(t)).
\end{equation}
Under the initial condition $f(0) = n$, the time-dependent solution reads
\begin{equation}
    f(t)=\frac{\gamma_1n+\gamma_2m}{\gamma_1+\gamma_2}+\frac{\gamma_2(n-m)}{\gamma_1+\gamma_2}e^{-(\gamma_1+\gamma_2)t}.
\end{equation}
Combined with the steady-state Green's functions, we find that the relaxation dynamics of Model 3 are identical to those of Model 1, as is also illustrated in Fig.~1(a) and 1(b).

\section{B. Out-of-time-ordered correlation functions}\label{OTOCs}
In this section, we investigate the out-of-time-ordered commutators and correlators (OTOCs). Our main focus is on the early- and late-time behaviors of these OTOCs, which provide measures to many-body quantum chaos. For the first two models, we provide a detailed evaluation of the early-time growth of the OTOCs and extract the corresponding quantum Lyapunov exponent. For Model 3, we demonstrate that the full quantum dynamics of the OTOCs can be solved exactly, and present its quantum Lyapunov exponent and the second freeness distance.

\subsection{Model 1: CBSYK(4) with dissipation coupling}\label{OTOCs1}
In this subsection, we calculate the early-time behavior and the quantum Lyapunov exponent of the OTOCs:
 \begin{equation}
  \begin{aligned}
  F_0(t)&=\text{tr}[\rho_{\mathrm{s.s.}}\{c_j^\dagger(0),c_k(t)\}\{c_j(0),c_k^\dagger(t)\}],\\
  F_1(t)&=-\text{tr}[\rho_{\mathrm{s.s.}}\{c_j^\dagger(0),c_k^\dagger(t)\}\{c_j(0),c_k(t)\}].
  \end{aligned}
  \end{equation}
To this end, we introduce the doubled Keldysh contour. The respective contributions to $F_0$ and $F_1$ can be expressed by the following contour diagrams:
\begin{center}
\begin{tikzpicture}[
    arrow inside/.style={
        postaction={decorate},
        decoration={
            markings,
            mark=at position 0.55 with {\arrow{Stealth[scale=1.2]}}
        }
    }
]
    \draw[thick] (-0.3, 3.9) -- (0.3, 3.9);
    \draw[thick]  (0, 3.9) -- (0, 3.2);
    
    \draw[thick, arrow inside] (0, 3.2) -- (4.3, 3.2);
    \draw[thick] (4.3, 3.2) .. controls (4.6, 3.2) and (4.6, 2.8) .. (4.3, 2.8);
    \draw[thick, arrow inside] (4.3, 2.8) -- (0, 2.8);
    
    \draw[thick] (0, 2.8) -- (0, 2.0);
    
    \draw[thick, arrow inside] (0, 2.0) -- (4.3, 2.0);
    \draw[thick] (4.3, 2.0) .. controls (4.6, 2.0) and (4.6, 1.6) .. (4.3, 1.6);
    \draw[thick, arrow inside] (4.3, 1.6) -- (0, 1.6);
    
    \draw[thick] (0, 1.6) -- (0, 1.0);
    
    \draw[thick] (0, 0.65) circle (0.35) node {$\rho_{ss}$};
    
    \draw[thick] (0, 0.3) -- (0, -0.2);
    \draw[thick] (-0.3, -0.2) -- (0.3, -0.2);
    
    \node[right] at (4.7, 3.2) {1};
    \node[right] at (4.7, 2.8) {2};
    \node[right] at (4.7, 2.0) {3};
    \node[right] at (4.7, 1.6) {4};
    
    \fill[black] (4.3, 3.2) circle (2pt);
    \fill[black] (4.3, 2.0) circle (2pt);
    \fill[red] (0, 3.2) circle (2pt);
    \fill[red] (0, 2.0) circle (2pt);
    \fill[blue] (0, 2.8) circle (2pt);
    \fill[blue] (0, 1.6) circle (2pt);
    
    \node[above=2pt] at (4.3, 3.2) {$\overline{\psi}_k$};
    \node[above=2pt] at (4.3, 2.0) {$\psi_k$};
    \node[left] at (-0.1, 3.0) {$\psi_j$};
    \node[left] at (-0.1, 1.8) {$\overline{\psi}_j$};

    \begin{scope}[xshift=7.2cm]
        \draw[thick] (-0.3, 3.9) -- (0.3, 3.9);
        \draw[thick]  (0, 3.9) -- (0, 3.2);
        
        \draw[thick, arrow inside] (0, 3.2) -- (4.3, 3.2);
        \draw[thick] (4.3, 3.2) .. controls (4.6, 3.2) and (4.6, 2.8) .. (4.3, 2.8);
        \draw[thick, arrow inside] (4.3, 2.8) -- (0, 2.8);
        
        \draw[thick] (0, 2.8) -- (0, 2.0);
        
        \draw[thick, arrow inside] (0, 2.0) -- (4.3, 2.0);
        \draw[thick] (4.3, 2.0) .. controls (4.6, 2.0) and (4.6, 1.6) .. (4.3, 1.6);
        \draw[thick, arrow inside] (4.3, 1.6) -- (0, 1.6);
        
        \draw[thick] (0, 1.6) -- (0, 1.0);
        
        \draw[thick] (0, 0.65) circle (0.35) node {$\rho_{ss}$};
        
        \draw[thick] (0, 0.3) -- (0, -0.2);
        \draw[thick] (-0.3, -0.2) -- (0.3, -0.2);
        
        \node[right] at (4.7, 3.2) {1};
        \node[right] at (4.7, 2.8) {2};
        \node[right] at (4.7, 2.0) {3};
        \node[right] at (4.7, 1.6) {4};
        
        \fill[black] (4.3, 3.2) circle (2pt);
        \fill[black] (4.3, 2.0) circle (2pt);
        \fill[red] (0, 3.2) circle (2pt);
        \fill[red] (0, 2.0) circle (2pt);
        \fill[blue] (0, 2.8) circle (2pt);
        \fill[blue] (0, 1.6) circle (2pt);
        
        \node[above=2pt] at (4.3, 3.2) {$\psi_k$};
        \node[above=2pt] at (4.3, 2.0) {$\overline{\psi}_k$};
        \node[left] at (-0.1, 3.0) {$\psi_j$};
        \node[left] at (-0.1, 1.8) {$\overline{\psi}_j$};
    \end{scope}
\end{tikzpicture}
\end{center}
The two red and blue dots signify the insertion of two operators, $\psi_j$ and $\bar{\psi}_j$, at time $t = 0$, while the two black dots denote the insertion of $\psi_k (\bar{\psi}_k)$ and $\bar{\psi}_k (\psi_k)$ at time $t$. For convenience, we label the fields and Green's functions according to their contour indices: $G^{ab}(t_1,t_2) = -i\langle \psi_i^a(t_1)\bar{\psi}_i^b(t_2)\rangle$, where $a,b \in \{1,2,3,4\}$. Adopting this convention, the full $4\times4$ Green's function and self-energy matrices are structured as:
\begin{equation}\label{eq:doubleGreensfun}
\hat{G}_{4\times4}(t_1,t_2) = 
\begin{pmatrix}
\hat{G}_{11}(t_1,t_2) & \hat{G}_{12}(t_1,t_2)  \\
\hat{G}_{21}(t_1,t_2) & \hat{G}_{22}(t_1,t_2) \\
\end{pmatrix}
,~~~
\hat{\Sigma}_{4\times4}(t_1,t_2) = 
\begin{pmatrix}
\hat{\Sigma}_{11}(t_1,t_2) & \hat{\Sigma}_{12}(t_1,t_2)  \\
\hat{\Sigma}_{21}(t_1,t_2) & \hat{\Sigma}_{22}(t_1,t_2) \\
\end{pmatrix}.
\end{equation}
Every block within $\hat{G}$ and $\hat{\Sigma}$ constitutes a $2 \times 2$ matrix. The diagonal blocks represent the single-contour Green's functions, which are well captured by the traditional Schwinger-Keldysh formalism, whereas the off-diagonal blocks denote the inter-contour Green's functions emerging from the doubled Keldysh contour structure. For this fermionic system, it is convenient to employ the RKA formalism, which corresponds to rotating the basis from $\psi^\pm$ to $\psi^{c/q}$ and $\bar{\psi}^\pm$ to $\bar{\psi}^{q/c}$ via:
\begin{equation}\label{Keldyshrot}
\begin{pmatrix}
\psi^c   \\
\psi^q  \\
\end{pmatrix}=\frac{1}{\sqrt{2}}\begin{pmatrix}
1 & 1  \\
1 & -1 \\
\end{pmatrix}
\begin{pmatrix}
\psi^+   \\
\psi^-  \\
\end{pmatrix},~~~
\begin{pmatrix}
\bar{\psi}^c   \\
\bar{\psi}^q  \\
\end{pmatrix}=\frac{1}{\sqrt{2}}\begin{pmatrix}
1 & 1  \\
1 & -1 \\
\end{pmatrix}
\begin{pmatrix}
\overline{\psi}^+   \\
\overline{\psi}^-  \\
\end{pmatrix}.
\end{equation}
In this rotated basis, the matrices take the form:
\begin{equation}\label{eq:rotdoubleGreensfun}
\overline{G}_{4\times4}(t_1,t_2) = 
\begin{pmatrix}
\overline{G}_{11}(t_1,t_2) & \overline{G}_{12}(t_1,t_2)  \\
\overline{G}_{21}(t_1,t_2) & \overline{G}_{22}(t_1,t_2) \\
\end{pmatrix}
,~~~
\overline{\Sigma}_{4\times4}(t_1,t_2) = 
\begin{pmatrix}
\overline{\Sigma}_{11}(t_1,t_2) & \overline{\Sigma}_{12}(t_1,t_2)  \\
\overline{\Sigma}_{21}(t_1,t_2) & \overline{\Sigma}_{22}(t_1,t_2) \\
\end{pmatrix},
\end{equation}
where the constitutive block matrices possess the following universal structures:
\begin{equation}\label{eq:blockmatrix}
\overline{G}_{11}= \overline{G}_{22}=
\begin{pmatrix}
G^R & G^K  \\
G^A & 0 \\
\end{pmatrix}
,~~~
\overline{G}_{12}= \overline{G}_{21}=
\begin{pmatrix}
0 & 2G^{W}  \\
0 & 0 \\
\end{pmatrix},
\end{equation}
and a similar relation holds for the self-energies. Here, we have introduced the Wightman Green's function $G^W_{12/21}(t_1,t_2) = G^{14/41}(t_1,t_2)$. Substituting these definitions into the Schwinger-Dyson equation yields:
\begin{equation}
\begin{pmatrix}
(\overline{G^0}_{11} )^{-1}-\overline{\Sigma}_{11}& -\overline{\Sigma}_{12}  \\
-\overline{\Sigma}_{21} & (\overline{G^0}_{22} )^{-1}-\overline{\Sigma}_{22} \\
\end{pmatrix}\circ
\begin{pmatrix}
\overline{G}_{11} & \overline{G}_{12}  \\
\overline{G}_{21} & \overline{G}_{22} \\
\end{pmatrix}=\mathds{1}_{4\times4}.
\end{equation}
Consequently, we obtain the following integral equations:
\begin{equation}\label{eq:RAWSDeq}
\begin{aligned}
 &\int dt_3dt_4G^R(t_{13})\Sigma^{14}(t_3,t_4)G^A(t_{42})=-G^{14}(t_1,t_2),\\
 &\int dt_3dt_4G^R(t_{13})\Sigma^{41}(t_3,t_4)G^A(t_{42})=-G^{41}(t_1,t_2).
\end{aligned}
\end{equation}
To express OTOCs using the doubled Keldysh contour, we view the operators inserted at $t=0$ as perturbations~\cite{Aleiner:2016eni,Gu:2021xaj}, and the operators at time $t$ then become a two-point function on the perturbed background.
\begin{equation}\label{eq:source}
\begin{aligned}
 \delta S=s_2\sum_i\left(\overline{\psi}^i_3(t_0)-\overline{\psi}^i_4(t_0)\right) \left(\psi^i_1(t_0)-\psi^i_2(t_0)\right)+s_1\sum_i\left(\psi^i_3(t_0)-\psi^i_4(t_0)\right) \left(\overline{\psi}^i_1(t_0)-\overline{\psi}^i_2(t_0)\right).
\end{aligned}
\end{equation}
The sources do not affect single-contour observables, and just contributes an additional self-energy term.
\begin{equation}\label{eq:perself31}
\begin{aligned}
        \Sigma^{41}(t_1,t_2)=&J\delta(t_1-t_2)\left(-G^{14}(t_2,t_1)\right)G^{41}(t_1,t_2)^2
    +V\delta(t_1-t_2)G^{41}_1(t_1,t_2)^2\left(-G^{14}_1(t_2,t_1)\right)\\
    &+\kappa\delta(t_1-t_2)\left(-G^{14}_2(t_2,t_1)\right)G^{41}_2(t_1,t_2)^2-is_2\delta(t-t_0)\delta(t'-t_0),
\end{aligned}
\end{equation}
\begin{equation}\label{eq:perself13}
\begin{aligned}
        \Sigma^{14}(t_1,t_2)=&J\delta(t_1-t_2)\left(-G^{41}(t_2,t_1)\right)G^{14}(t_1,t_2)^2
    +V\delta(t_1-t_2)G^{14}_1(t_1,t_2)^2\left(-G^{41}_1(t_2,t_1)\right)\\
    &+\kappa\delta(t_1-t_2)\left(-G^{41}_2(t_2,t_1)\right)G^{14}_2(t_1,t_2)^2+is_1\delta(t-t_0)\delta(t'-t_0).
\end{aligned}
\end{equation}
Utilizing Eq.~(\ref{eq:RAWSDeq}) along with the explicit forms of the retarded and advanced Green's functions, we arrive at the differential equations:
\begin{equation}\label{eq:Gdfeq}
    \left(\partial_{t_1}+\Gamma/2 \right)\left(\partial_{t_2}+\Gamma/2 \right)G^{w\bar{w}}(t_1,t_2)=\Sigma^{w\bar{w}}(t_1,t_2).
\end{equation}
Here, we define $\overline{w} \neq w \in \{1,4\}$.
The right-hand sides of these equations only contains delta functions, which separate the solution into different regions as in the Majorana Brownian SYK case~\cite{Gu:2021xaj,Zhangp_2023}. Since the equations (and initial conditions) are symmetric with respect to $t_1$ and $t_2$, we assume:
\begin{equation}
\left\{
\begin{aligned}
    &G^{w\overline{w}} = e^{-\frac{\Gamma}{2}t_1} f_{w\overline{w}}(t_2) + e^{-\frac{\Gamma}{2}t_2} g_{w\overline{w}}(t_1), \quad &&(t_1, t_2) \in A, \\
    &G^{w\overline{w}} = e^{-\frac{\Gamma}{2}t_2} f_{w\overline{w}}(t_1) + e^{-\frac{\Gamma}{2}t_1} g_{w\overline{w}}(t_2), \quad &&(t_1, t_2) \in B, \\
    &G^{14}=-fe^{-\frac{\Gamma}{2}|t_{12}|},~G^{41}=(1-f)e^{-\frac{\Gamma}{2}|t_{12}|}, \quad &&(t_1, t_2) \in C \cup D, \\
    &G^{w\overline{w}}(t_0^+, t_0^+) = G^{w\overline{w}}(t_0^-, t_0^-) +i (-1)^w s_w.
\end{aligned}
\right.
\qquad\qquad 
\vcenter{\hbox{ 
\begin{tikzpicture}[scale=1.1, >=stealth]
    \fill[green] (0,0) -- (1,1) -- (0,1) -- cycle;
    \fill[green] (0,0) -- (1,0) -- (1,1) -- cycle;
    
    \draw[->] (-0.8,0) -- (1.4,0) node[right] {$t_1$};
    \draw[->] (0,-0.6) -- (0,1.4) node[above] {$t_2$};
    
    \draw[thick] (0,0) rectangle (1,1);
    
    \draw[thick, red] (-0.5,-0.5) -- (1.2,1.2);
    
    \node at (0.3, 0.7) {$A$};
    \node at (0.7, 0.3) {$B$};
    
    \node[left] at (-0.1, 0.5) {$C$};
    \node[below] at (0.6, -0.1) {$D$};
    \node[below right] at (0, 0) {$t_0$};
\end{tikzpicture}
}}
\label{eq:regions}
\end{equation}
For the third line, we use the fact that the source~(\ref{eq:source}) can be neglected due to the cancellation between $u$ and $d$ branches for either $t_1 < t_0$ or $t_2 < t_0$. For the fourth line, we integrate the equation (16) over a small square surrounding $(t_0, t_0)$ and use the continuum condition without the source term $G^W_{w\overline{w}}(t_0^+, t_0^-) = G^W_{w\overline{w}}(t_0^-, t_0^+) = G^W_{w\overline{w}}(t_0^-, t_0^-)$.

To determine the solution of $f_{w\overline{w}}$ and $g_{w\overline{w}}$, we match the boundary condition near the $AC$ and $AB$ boundary. Near the $AC$ boundary we have
\begin{equation}
(\partial_{t_2} + \frac{\Gamma}{2})f_{w\overline{w}}(t_2) = 0, \qquad f_{w\overline{w}} = a e^{-\frac{\Gamma t_2}{2}}.
\label{eq:boundary_ac}
\end{equation}
We could always fix $a = 0$ using the redundancy of $(f_{w\overline{w}}(t), g_{w\overline{w}}(t)) \rightarrow (f_{w\overline{w}}(t) - a e^{-\frac{\Gamma t}{2}}, g_{w\overline{w}}(t) + a e^{-\frac{\Gamma t}{2}})$. The boundary condition near $AB$ gives
\begin{equation}\label{eq:J4k31}
    \left(\partial_t+\Gamma\right)z_{w\overline{w}}(t)=\tilde\gamma_0z_{w\overline{w}}^2z_{\overline{w}w}+\gamma_1\frac{G_1^{w\overline{w}}}{G^{w\overline{w}}}+\gamma_2\frac{G_2^{w\overline{w}}}{G^{w\overline{w}}},
\end{equation}
where we have introduced the shorthand notations $G^{w\overline{w}} \equiv G^{w\overline{w}}(t_0^-, t_0^-)$, $G_{1/2}^{14} = in(m)$, and $G_{1/2}^{41} =-i(1 - n(m))$. $z_{w\overline{w}}=z_{\overline{w}w}=1$ is the unstable fixed point of~Eq.(\ref{eq:J4k31}), and the equation satisfied by the OTOCs before the scrambling time is given by the linearization of~Eq.(\ref{eq:J4k31}):
\begin{equation}\label{eq:J4k31linearize}
    \partial_t\delta z_{w\overline{w}}=(\tilde\gamma_0-\gamma_1-\gamma_2)\delta z_{w\overline{w}}+\tilde\gamma_0\delta z_{\overline{w}w}+\text{o}(\delta z^2).
\end{equation}
After identifying $F_{0/1}(t) = iG^{41/14}\delta z_{41/14}(t)$, we arrive at Eq.~(10) in the main text.

\subsection{Model 2: CBSYK(4) without dissipation coupling}\label{OTOCs2}
For our second model, the only difference from the first model lies in the explicit form of the self-energies, which are given by:
\begin{equation}\label{eq:perself231}
\begin{aligned}
        \Sigma^{41}(t_1,t_2)=&J\delta(t_1-t_2)\left(-G^{14}(t_2,t_1)\right)G^{41}(t_1,t_2)^2
    +V\delta(t_1-t_2)G^{41}_1(t_1,t_2)^2\left(-G^{14}_1(t_2,t_1)\right)\\
    &+\kappa\delta(t_1-t_2)\left(-G^{14}(t_2,t_1)\right)G^{41}_2(t_1,t_2)^2-is_2\delta(t-t_0)\delta(t'-t_0),
\end{aligned}
\end{equation}
\begin{equation}\label{eq:perself213}
\begin{aligned}
        \Sigma^{14}(t_1,t_2)=&J\delta(t_1-t_2)\left(-G^{41}(t_2,t_1)\right)G^{14}(t_1,t_2)^2
    +V\delta(t_1-t_2)G^{14}_1(t_1,t_2)^2\left(-G^{41}_1(t_2,t_1)\right)\\
    &+\kappa\delta(t_1-t_2)\left(-G^{41}(t_2,t_1)\right)G^{14}_2(t_1,t_2)^2+is_1\delta(t-t_0)\delta(t'-t_0).
\end{aligned}
\end{equation}
Following a similar procedure to that outlined in the preceding section, we extract the differential equations governing the perturbed two-point functions and the OTOCs:
\begin{equation}\label{eq:J4k22}
    \left(\partial_t+\Gamma\right)z_{w\overline{w}}(t)=\tilde\gamma_0z_{w\overline{w}}^2z_{\overline{w}w}+\gamma_1\frac{G_1^{w\overline{w}}}{G^{w\overline{w}}}+\gamma_2\frac{G^{\overline{w}w}G_2^{w\overline{w}}}{G^{w\overline{w}}G_2^{\overline{w}w}}z_{\overline{w}w}.
\end{equation}
\begin{equation}\label{eq:J4k22linearize}
    \partial_t\bm{F}(t)=(\tilde\gamma_0-\gamma_1-\gamma_2)\bm{F}(t)+\left(\tilde\gamma_0+\gamma_2\frac{G^{\overline{w}w}G_2^{w\overline{w}}}{G^{w\overline{w}}G_2^{\overline{w}w}}\right)\sigma_x\bm{F}(t).
\end{equation}
The quantum Lyapunov exponent $\varkappa$ is identified with the largest eigenvalue of the right-hand side, yielding
  \begin{equation}\label{eq:Lyap2}
\varkappa=\sqrt{(\tilde\gamma_0+\gamma_2)^2+J\kappa(m-f)^2}+\left(1-\frac{\kappa}{J}\right)\tilde\gamma_0-\kappa(m-f)^2-\gamma_1.
  \end{equation}

\subsection{Model 3: CBSYK(3,1) with dissipation coupling}\label{OTOCs3}
For our third model, the corresponding self-energies are explicitly evaluated as:
\begin{equation}\label{eq:perself331}
\begin{aligned}
        \Sigma^{41}&(t_1,t_2)=\frac{J}{2}\delta(t_1-t_2)\Big(\left(-G^{14}(t_2,t_1)\right)G^{41}(t_1,t_2)^2+\left(-G^{14}(t_2,t_1)\right)G^{41}(t_1,t_2)G_1^{41}(t_1,t_2)\Big)\\
    &~~+V\delta(t_1-t_2)G^{41}_1(t_1,t_2)^2\left(-G^{14}_1(t_2,t_1)\right)
    +\kappa\delta(t_1-t_2)\left(-G^{14}(t_2,t_1)\right)G^{41}_2(t_1,t_2)^2-is_2\delta(t-t_0)\delta(t'-t_0),
\end{aligned}
\end{equation}
\begin{equation}\label{eq:perself313}
\begin{aligned}
        \Sigma^{14}&(t_1,t_2)=\frac{J}{2}\delta(t_1-t_2)\Big(\left(-G^{41}(t_2,t_1)\right)G^{14}(t_1,t_2)^2+\left(-G^{41}(t_2,t_1)\right)G^{14}(t_1,t_2)G_1^{14}(t_1,t_2)\Big)\\
    &~~+V\delta(t_1-t_2)G^{14}_1(t_1,t_2)^2\left(-G^{41}_1(t_2,t_1)\right)
    +\kappa\delta(t_1-t_2)\left(-G^{41}(t_2,t_1)\right)G^{14}_2(t_1,t_2)^2+is_1\delta(t-t_0)\delta(t'-t_0).
\end{aligned}
\end{equation}
Following the same approach as detailed above, the equations governing the perturbed two-point functions and the OTOCs reduce to:
\begin{equation}\label{eq:J31k31}
    \left(\partial_t+\Gamma\right)z_{w\bar{w}}(t)=\tilde\gamma_0z_{w\bar{w}}^2z_{\bar{w}w}+\gamma_1\frac{G_1^{w\bar{w}}}{G^{w\bar{w}}}+\gamma_2\frac{G_2^{w\bar{w}}}{G^{w\bar{w}}},
\end{equation}
\begin{equation}\label{eq:J31k31linearize}
    \partial_t\bm{F}(t)=(\tilde\gamma_0-\gamma_1-\gamma_2)\bm{F}(t)+\tilde\gamma_0\sigma_x\bm{F}(t).
\end{equation}
The quantum Lyapunov exponent $\varkappa$ is identified with the largest eigenvalue of the right-hand side, yielding
\begin{equation}\label{eq:Lyap3}
  \varkappa=2\tilde{\gamma}_0-\gamma_1-\gamma_2=\varkappa_0+2(\tilde{\gamma}_0-\gamma_0)-\gamma_2,
  \end{equation}
which is identically to the result obtained for Model 1.

\subsubsection{2-Freeness}
Freeness, which characterizes the statistical independence within a non-commutative probability space~\cite{nica2006lectures,speicher2009freeprobabilitytheory,mingo2017free}, has recently been proposed as a diagnostic tool for quantum chaos in many-body systems~\cite{Fava_2025, camargo2025quantumsignatureschaosfree}. Our objective is to elucidate whether this framework yields qualitatively identical conclusions to those established by the quantum Lyapunov exponent for the systems with sufficiently long range interactions. Chaotic Hamiltonian evolution is argued to drive operators toward asymptotic freeness~\cite{Fava_2025}, which is manifested by the vanishing
of all mixed free cumulants. The mixed free cumulants vanish up to order of 2k is called k-freeness~\cite{Fava_2025}. In what follows, we will demonstrate that within the present models, the late-time behavior of the OTOCs is identical to the fourth free cumulant. We then evaluate the late-time residual values of OTOCs, which serve as a measure of the 2-freeness.

We begin by briefly reviewing the concept of free cumulants and proving that the long-time limit of the OTOC coincides with the fourth free cumulant. Free cumulants are defined implicitly through the combinatorics of non-crossing partitions as~\cite{nica2006lectures,speicher2009freeprobabilitytheory,mingo2017free}:
\begin{equation}\label{eq:impfreecum}
\langle\hat{A}_{n-1}(t_{n-1})\hat{A}_{n-2}(t_{n-2})\cdots\hat{A}_1(t_1)\hat{A}_0(t_{0})\rangle=\sum_{\pi\in NC(n)}\kappa_\pi(t_{n-1},\cdots,t_0).
\end{equation}
For the charged operator, do not preserve particle number, $\langle \hat{c}_i \rangle=\langle \hat{c}_i ^\dagger\rangle=0$. Consequently, the first four free cumulants reduce to:
\begin{equation}\label{eq:freecum}
\begin{aligned}
    &~~~~~~\kappa_1(0)=\kappa_3(t,0,t)=0,~~~~~\kappa_2(t,0)=\langle\hat{c}_i(t)\hat{c}_j^\dagger(0)\rangle,\\
    &\kappa_4(t,0,t,0)=\langle\hat{c}_i(t)\hat{c}_j(0)\hat{c}_i^\dagger(t)\hat{c}_j^\dagger(0)\rangle-\langle\hat{c}_i(t)\hat{c}_j^\dagger(0)\rangle \langle\hat{c}_j(0)\hat{c}_i^\dagger(t)\rangle,
\end{aligned}
\end{equation}
Recalling that the steady-state two-point Green's functions~Eq.(\ref{eq:Boltzt}) decay to zero asymptotically in the long-time limit, we immediately conclude that the late-time OTOCs are asymptotically equivalent to the fourth free cumulants. In the following subsections, we evaluate the late-time behavior of these OTOCs to diagnose the 2-freeness of the system.
\subsubsection{Scramblon Effective Theory}
In the late-time regime, OTOCs are mediated by collective modes known as scramblons~\cite{Kitaev_2018a,Gu:2021xaj}. In this subsection, we first extract the scramblon functions for the CBSYK model and then utilize the scramblon effective theory to evaluate the full dynamics of OTOCs.

\begin{figure}[htbp]
    \centering
    \includegraphics[width=0.95\linewidth]{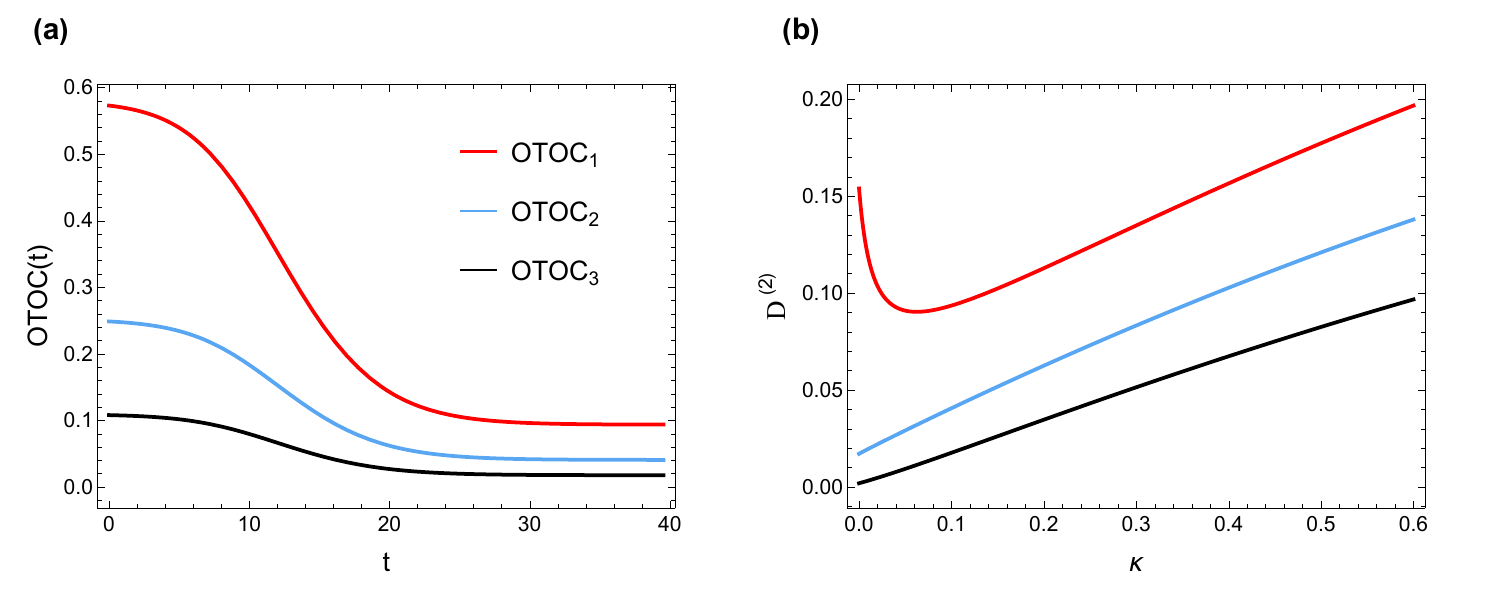}
    \caption{(a) Full dynamics of the three OTOCs as a function of t for a fixed coupling $\kappa = 0.1$. (b) Freeness distances of the three OTOCs as a function of the coupling strength $\kappa$. All curves are obtained from Model 3 with parameters: $J=1$, $V = 0.1$, $m = 0.45$, and $n = 0.1$.}
    \label{fig:OTOC}
  \end{figure}

We start by solving Eq.(\ref{eq:J31k31}), with the sources: $s_1=fs$, $s_2=(1-f)s$,
\begin{equation}\label{3131earlyz}
    z_{w\overline{w}}(t)=r+\frac{1-r}{1+z_{w\overline{w}}e^{\varkappa(t-t_0)}}.
\end{equation}
Here, $\varkappa=2\tilde\gamma_0-\gamma_1-\gamma_2$, $r=\frac{\gamma_1+\gamma_2}{2\tilde\gamma_0}$, and z is determined by the initial condition: $z_{14/41}(t_0)=1-s_{1/2}$.
\begin{equation}\label{3131earlyG}
\begin{aligned}
    iG&^{41}(t_1,t_2)=(1-f)e^{-\Gamma|t_{12}|/2}\left(r+\frac{1-r}{1+\frac{p}{1-r}e^{\varkappa\frac{t_1+t_2-|t_{12}|}{2}}}\right)\\
    &=\sum_{m,n=0}^\infty\frac{1}{m!n!}\Upsilon_{41}^{R,m+n}(t_{12})\left[-\frac{e^{\varkappa\frac{t_1+t_2}{2}}}{C}\right]^{m+n}\left[2Np(1-f)\Upsilon_{41}^{A,1}(0)\right]^m\left[-2Npf\Upsilon_{14}^{A,1}(0)\right]^n\\
    &=\sum_{k,m=0}^\infty\frac{\binom{k}{m}}{k!}\Upsilon_{41}^{R,k}(t_{12})\left[-\frac{e^{\varkappa\frac{t_1+t_2}{2}}}{C}\right]^{k}\left[4Np(1-f)\Upsilon_{41}^{A,1}(0)\right]^k\\
    &=\sum_{k=0}^\infty\frac{1}{k!}\Upsilon_{41}^{R,k}(t_{12})\left[-\frac{e^{\varkappa\frac{t_1+t_2}{2}}}{C}\right]^{k}\left[4Np(1-f)\Upsilon_{41}^{A,1}(0)\right]^k.
\end{aligned}
\end{equation}
In the first line, we set $s=pe^{\varkappa t_0}$, with $t_0\rightarrow-\infty$ to probe the distribution of scramblons perturbations. The second line is given by the scramblon effective theory~\cite{Gu:2021xaj}, where $-\frac{e^{\varkappa\frac{t_1+t_2}{2}}}{C}$ is the propagator of the fastest growing mode: scramblon, the factor 2 comes from two possible OTOCs generated by each source term, and the factor N is due to the summation over indices in Eq.(\ref{eq:source}). By identifying $\Upsilon_{14}^{A,1}=-\frac{1-f}{f}\Upsilon_{41}^{A,1}$, and use the formula $\frac{1}{m!n!}=\frac{\binom{m+n}{m}}{(m+n)!}$ in the third line,
We extract the scramblon functions:
\begin{equation}\label{feymanrule1}
\begin{aligned}
\Upsilon_{41}^{R/A,0}(t)=(1-f)e^{-\Gamma|t|/2}
,~~~\Upsilon_{41}^{R/A,k\neq0}(t)=(1-r)(1-f)k!e^{-(\Gamma+k\varkappa)|t|/2},
\end{aligned}    
\end{equation}

\begin{equation}\label{feymanrule2}
\begin{aligned}
\Upsilon_{14}^{R/A,0}(t)=-fe^{-\Gamma|t|/2}
,~~~\Upsilon_{14}^{R/A,k\neq0}(t)=-(1-r)fk!e^{-(\Gamma+k\varkappa)|t|/2},
\end{aligned}    
\end{equation}

\begin{equation}\label{feymanrule3}
C=4N(1-r)^2f(1-f),
\end{equation}
We present the basic Feynman diagrams in the Scramblon effective theory~\cite{Gu:2021xaj} as follows: 
\begin{equation}
G^{w\overline{w}}(t_1, t_2) = 
\begin{tikzpicture}[baseline=(current bounding box.center), scale=1.2]
    \tikzset{
        blob/.style={circle, fill=black!20, draw=black, thick, minimum size=6mm, inner sep=0pt},
        psphere1/.style={circle, fill=black!20, draw=black, thick, minimum size=4mm, inner sep=0pt, font=\scriptsize},
        psphere2/.style={circle, fill=cyan!30, draw=black, thick, minimum size=4mm, inner sep=0pt, font=\scriptsize},
        propagator/.style={thick, decorate, decoration={snake, amplitude=1.0pt, segment length=6.0pt}}
    }

    \node[blob, label={[above, yshift=0.5mm]:$\Upsilon^{R,m+n}_{w\overline{w}}$}] (C) at (0,0) {};
    
    \draw[thick] (C) -- (-0.65, 0.45) node[left] {$t_1$};
    \draw[thick] (C) -- (-0.65, -0.45) node[left] {$t_2$};
    
    \node[psphere1, label={[above, yshift=1mm]:$\Upsilon^{A,1}_{w\overline{w}}$}] (P1) at (42:1.1) {$p$};
    \node[psphere1] (P2) at (0:1.1) {$p$};
    \node[psphere2, label={[below left, xshift=-1mm, yshift=1mm]:$\Upsilon^{A,1}_{\overline{w}w}$}] (P3) at (-42:1.1) {$p$};
    
    \draw[propagator] (C) -- (P1);
    \draw[propagator] (C) -- (P2);
    \draw[propagator] (C) -- (P3);
    
    \draw[thick] (P1) -- ++(0, 0.32);
    \draw[thick] (P1) -- ++(0.32, 0);
    
    \draw[thick] (P2) -- ++(0.26, 0.22);
    \draw[thick] (P2) -- ++(0.26, -0.22);
    
    \draw[thick] (P3) -- ++(0, -0.32);
    \draw[thick] (P3) -- ++(0.32, 0);
\end{tikzpicture}
\quad , \quad
\text{OTOC}(t_1, t_2; t_3, t_4) = 
\begin{tikzpicture}[baseline=(current bounding box.center), scale=1.2]
    \tikzset{
        blob/.style={circle, fill=black!20, draw=black, thick, minimum size=6mm, inner sep=0pt},
        propagator/.style={thick, decorate, decoration={snake, amplitude=1.0pt, segment length=6.0pt}}
    }

    \node[blob, label={[above, yshift=0.6mm]:$\Upsilon^{R,n}$}] (L) at (0,0) {};
    \node[blob, label={[above, yshift=0.6mm]:$\Upsilon^{A,n}$}] (R) at (1.5,0) {};
    
    \draw[thick] (L) -- (-0.8, 0.5) node[left] {$t_1$};
    \draw[thick] (L) -- (-0.8, -0.5) node[left] {$t_2$};
    
    \draw[thick] (R) -- (2.3, 0.5) node[right] {$t_3$};
    \draw[thick] (R) -- (2.3, -0.5) node[right] {$t_4$};
    
    \draw[propagator] (L) to[bend left=25] (R);
    \draw[propagator] (L) to[bend left=7] (R);
    \draw[propagator] (L) to[bend right=7] (R);
    \draw[propagator] (L) to[bend right=25] (R);
\end{tikzpicture}
\end{equation}
The balls and wave lines represent vortices and the propagating scramblons modes, respectively. The three different OTOCs are then given by summing all scramblons contributions:
\begin{equation}\label{OTOC1}
\begin{aligned}
    \mathrm{OTOC}_1(t_1,t_2;&t_3,t_4)=-\frac{1}{N^2}\sum_{ij}\langle\hat{c}_i(t_1)\hat{c}_j(t_3)\hat{c}_i^\dagger(t_2)\hat{c}_j^\dagger(t_4)\rangle\\
    &=\sum_{k=0}^{\infty}\frac{(-\lambda)^k}{k!}\Upsilon^{R,k}_{41}(t_{12})\Upsilon^{A,k}_{41}(t_{34})\\
    &=D^{(2)}_1-G^{41}(t_{12})G^{41}(t_{34})\left[\frac{e^{\varkappa\frac{|t_{12}|+|t_{34}|}{2}}}{\lambda}\right]U\left(1,1,\frac{e^{\varkappa\frac{|t_{12}|+|t_{34}|}{2}}}{\lambda}\right),
\end{aligned}
\end{equation}

\begin{equation}\label{OTOC2}
\begin{aligned}
    \mathrm{OTOC}_2(t_1,t_2;&t_3,t_4)=-\frac{1}{N^2}\sum_{ij}\langle \hat{c}_i(t_1)\hat{c}^\dagger_j(t_3)\hat{c}_i^\dagger(t_2)\hat{c}_j(t_4)\rangle\\
    &=-\sum_{k=0}^{\infty}\frac{(-\lambda)^k}{k!}\Upsilon^{R,k}_{41}(t_{12})\Upsilon^{A,k}_{14}(t_{34})\\
    &=D^{(2)}_2-G^{41}(t_{12})G^{14}(t_{34})\left[\frac{e^{\varkappa\frac{|t_{12}|+|t_{34}|}{2}}}{\lambda}\right]U\left(1,1,\frac{e^{\varkappa\frac{|t_{12}|+|t_{34}|}{2}}}{\lambda}\right),
\end{aligned}
\end{equation}

\begin{equation}\label{OTOC3}
\begin{aligned}
    \mathrm{OTOC}_3(t_1,t_2;&t_3,t_4)=-\frac{1}{N^2}\sum_{ij}\langle\hat{c}^\dagger_i(t_1)\hat{c}^\dagger_j(t_3)\hat{c}_i(t_2)\hat{c}_j(t_4)\rangle\\
    &=\sum_{k=0}^{\infty}\frac{(-\lambda)^k}{k!}\Upsilon^{R,k}_{14}(t_{12})\Upsilon^{A,k}_{14}(t_{34})\\
    &=D^{(2)}_3-G^{14}(t_{12})G^{14}(t_{34})\left[\frac{e^{\varkappa\frac{|t_{12}|+|t_{34}|}{2}}}{\lambda}\right]U\left(1,1,\frac{e^{\varkappa\frac{|t_{12}|+|t_{34}|}{2}}}{\lambda}\right).
\end{aligned}
\end{equation}
Here, $-\lambda=-\frac{e^{\varkappa\frac{t_1+t_2-t_3-t_4}{2}}}{C}$ is the propagator of scramblon, $k!$ is the symmetry factor. $D_1^{(2)}=(1-f)^2r(2-r)$, $D_2^{(2)}=f(1-f)r(2-r)$, $G^{14}(t_{12})=fe^{-\frac{\Gamma}{2}|t_{12}|}$, $G^{41}(t_{12})=(1-f)e^{-\frac{\Gamma}{2}|t_{12}|}$ and $D_3^{(2)}=f^2r(2-r)$ are the second freeness distance. We present the corresponding numerical results in FIG.\ref{fig:OTOC}.

We find that for $\text{OTOC}_1$, a small $\kappa$ decreases the 2-freeness distance $D^{(2)}_1$, but $D^{(2)}_1$ monotonically increases with $\kappa$ in the larger $\kappa$ regime, which is consistent with the behavior of the quantum Lyapunov exponent. However, for $\text{OTOC}_2$ and $\text{OTOC}_3$, the 2-freeness distances monotonically increase with $\kappa$, even in the regime of small $\kappa$ and sufficiently large $m-n$ (see the discussion following Eq.(12) in the main text), exhibiting a clear departure from the Lyapunov exponent behavior. Therefore, we conclude that, in general, the quantum Lyapunov exponent and the freeness distance serve as independent measure for many-body quantum chaos that do not always yield qualitatively identical conclusions.

\end{document}